\documentclass[12pt]{spieman}  % 12pt font required by SPIE;
\usepackage{amsmath,amsfonts,amssymb}
\usepackage{graphicx}
\usepackage{setspace}
\usepackage{tocloft}
\usepackage{lineno}

\usepackage{booktabs} 

\usepackage[numbers,sort&compress]{natbib}         % or abbrvnat, unsrtnat

\linenumbers
\title{Design Optimized CMOS-Compatible Hybrid Silicon–MIM Waveguide Cavity for Label-Free Sensing}
\usepackage{algorithm}        % provides the floating 'algorithm' env
\usepackage{algpseudocode}    % provides the 'algorithmic' env and commands
\author[a]{Alireza~Khalilian}
\author[b]{Hima~Nikafshan~Rad}

\AtBeginDocument{\nolinenumbers}  % turn off globally
\affil[a]{University of Michigan-Dearborn, Integrated Nano Optoelectronics Laboratory, Evergreen Rd., Dearborn, Michigan 48128, USA}
\affil[b]{Griffith University, School of Information and Communication Technology (ICT), Brisbane, Australia}

\cftpagenumbersoff{figure}
\cftpagenumbersoff{table} 
\begin{document} 
\maketitle

\begin{abstract}
We numerically design a hybrid plasmonic–dielectric refractive-index biosensor comprising a metal–insulator–metal (MIM) bus waveguide evanescently coupled to a silicon-core resonator. The geometry is optimized with a genetic algorithm (GA) using Ansys Lumerical FDTD simulations, targeting maximal sensitivity and figure of merit while minimizing the resonance linewidth. The optimized device achieves a sensitivity of $1276\,\mathrm{nm/RIU}$, a full width at half maximum (FWHM) of $11.89\,\mathrm{nm}$, a quality factor of $221$, and a figure of merit of $107\,\mathrm{RIU}^{-1}$ (defined as $S/\mathrm{FWHM}$). These results demonstrate precise discrimination of small refractive-index variations and support the sensor’s applicability to label-free biochemical and medical diagnostics.
\end{abstract}

% Include a list of up to six keywords after the abstract
\keywords{Hybrid plasmonics, MIM waveguide, Silicon microring, RI biosensing}

% Include email contact information for corresponding author
{\noindent \footnotesize\textbf{*}Alireza Khalilian,  \linkable{akhalili@umich.edu} }

\begin{spacing}{2}   % use double spacing for rest of manuscript

\section{Introduction}
\label{sect:intro}  % \label{} allows reference to this section

Refractive-index (RI) sensing underpins a wide range of label-free assays in biomedicine, chemical analysis, and food quality control \citep{parizi2021high,ahmed2019refractive,poscio1990realization,khalilian2017highly}. In these platforms, a small change $\Delta n$ in the analyte’s RI manifests as a measurable spectral shift $\Delta\lambda$, with sensitivity defined as $S=\Delta\lambda/\Delta n$ (nm/RIU). Performance is commonly summarized by the full width at half maximum (FWHM) $\Gamma$, the quality factor $Q=\lambda_0/\Gamma$, and the figure of merit $\mathrm{FOM}=S/\Gamma$. Microring resonators are a natural fit for RI sensing because they support narrow resonances and steep dispersion, enabling precise spectral readout \citep{yi2016tunable,yi2017plasmonic,patel2022graphene}. High-$Q$ dielectric rings can detect minute perturbations in the surrounding medium \citep{gylfason2010chip,huang2021high}. Their principal limitation is overlap: most of the optical energy is confined to the core, and only the evanescent tail samples the analyte, which caps $S$ \citep{zhang2021all,maksimov2022enhanced}. A second practical constraint is miniaturization—bending loss grows rapidly as the radius shrinks, degrading $Q$ and, in turn, the FOM. Plasmonic waveguides address the overlap problem through subwavelength confinement and strong field localization at metal–dielectric interfaces \citep{jain2022photonic,pal2021theoretical}. This benefit, however, comes with Ohmic dissipation and shorter propagation lengths than their dielectric counterparts \citep{mahfuz2019bimetallic,ilchenko2017using}. Hybrid plasmonic waveguides (HPWs) strike a middle ground by combining a dielectric channel with a nearby metal layer to concentrate the field in the sensing region while tempering loss, and they can be realized in CMOS-compatible stacks \citep{guo2022integrated,ou2021wide,zhu2019temperature}. Recent researches illustrate the trade-space. Kumar \textit{et~al.} used a slotted HPW ring to intensify light–analyte interaction, reporting a $29.6\,\mathrm{nm}$ resonance shift and $S=1609\,\mathrm{nm/RIU}$ for polluted-water detection \citep{kumar2023nanophotonic}. Guo \textit{et~al.} demonstrated suspended chalcogenide slot HPW microrings with low propagation loss and $S=511.5\,\mathrm{nm/RIU}$ \citep{guo2023ultra}. Butt \textit{et~al.} realized a metal–air–silicon hybrid microring with sensitivities of $690\,\mathrm{nm/RIU}$ (gas) and $401\,\mathrm{nm/RIU}$ (biosensing) \citep{butt2020highly}. Together, these results underscore that high $S$ must be balanced against linewidth and propagation loss to maximize $\mathrm{FOM}$.

Here we develop a CMOS-compatible hybrid plasmonic resonator cavity for biochemical sensing that explicitly targets this balance. The cavity supports a hybrid mode engineered to (i) increase analyte–field overlap relative to a purely dielectric ring while (ii) maintaining a manageable linewidth. To navigate the competing objectives, we couple a genetic algorithm (GA) with Ansys Lumerical finite-difference time-domain (FDTD) simulations to tune the core shape and dimensions under fabrication-realistic constraints. The optimized design achieves high $S$ and $\mathrm{FOM}$ without sacrificing spectral clarity, providing a compact and integration-ready route to precise RI sensing.

\section{Proposed biosensing resonator cavity}

\begin{figure}[htbp]
\centering\includegraphics[width=13cm]{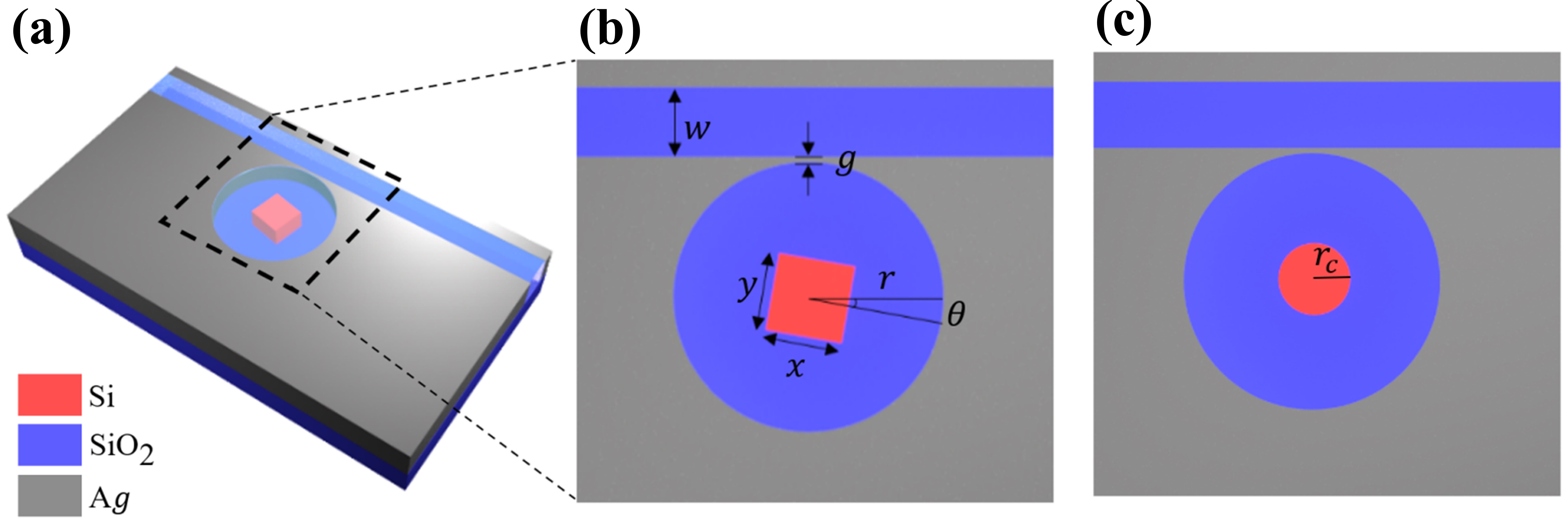}
\caption{(a) 3D view of the hybrid resonator cavity with a square silicon core. (b) Planar view and geometric parameters for the square-core design. (c) Planar view for the circular-core design.}
\label{fig:1}
\end{figure}

Figure~\ref{fig:1}(a) shows a three-dimensional schematic of the hybrid metal–insulator–metal (MIM) plasmonic sensor. The device consists of a MIM bus waveguide side-coupled to a ring resonator that contains a central silicon inclusion (square or circular). Planar $(x\text{–}y)$ cross-sections with geometric definitions are provided in Figure~\ref{fig:1}(b,c) for the square- and circular-core variants, respectively. Two configurations are considered: (i) a square silicon block at the cavity center, rotated by an angle $\theta$, and (ii) a circular silicon block. Unless noted otherwise, the ring outer radius is $r=1400\,\mathrm{nm}$, the waveguide–ring gap is $g=10\,\mathrm{nm}$, and the MIM bus width is $w=450\,\mathrm{nm}$. For the circular inclusion, $r_c$ denotes the silicon radius; for the square inclusion, the side lengths and rotation are $x=906\,\mathrm{nm}$, $y=854.2\,\mathrm{nm}$, and $\theta=67.25^\circ$. In the material map, silver (Ag) and air are labeled by $\varepsilon_m$ and $\varepsilon_i$, respectively.

Silver is selected due to its comparatively low absorption in the simulated band relative to Au and Al. Its dispersive permittivity is modeled with a Drude response \cite{di2014overview}:
\begin{equation}
\varepsilon_m(\omega)=\varepsilon_\infty-\frac{\omega_p^2}{\omega(\omega+i\gamma)},
\label{eq:Drude}
\end{equation}
with $\varepsilon_\infty=3.7$, $\omega_p=9.1~\mathrm{eV}$, and $\gamma=0.018~\mathrm{eV}$. Excitation is provided from the left port of the bus waveguide; transmitted power is recorded at the right port. The transmittance is computed as $T=P_{\text{out}}/P_{\text{in}}$ using Ansys Lumerical FDTD. With $w=450\,\mathrm{nm}$, only the fundamental transverse-magnetic mode ($\mathrm{TM}_0$) is supported in the MIM bus over the operating wavelengths considered.

When the resonance condition is satisfied, surface plasmon polaritons (SPPs) in the bus couple to the ring, producing a narrow spectral dip/peak in $T(\lambda)$ \cite{rakhshani2019refractive,zhang2019refractive,butt2021metal}. For reference, the effective index $n_{\mathrm{eff}}$ of the guided SPP mode is obtained by solving the standard MIM dispersion relation; in our notation
\begin{equation}
\frac{\varepsilon_i\,p}{\varepsilon_m\,k}
=
\frac{1 - e^{-k w}}{1 + e^{-k w}},
\qquad
p=k_0\sqrt{n_{\mathrm{eff}}^2-\varepsilon_i},\quad
k=k_0\sqrt{n_{\mathrm{eff}}^2-\varepsilon_m},
\label{eq:dispersion}
\end{equation}
and
\begin{equation}
\beta_{\mathrm{spp}}=n_{\mathrm{eff}}k_0=\frac{2\pi n_{\mathrm{eff}}}{\lambda},
\label{eq:beta}
\end{equation}
where $k_0=2\pi/\lambda$ is the free-space wavenumber, $\varepsilon_i$ and $\varepsilon_m$ are the dielectric and metal permittivities, $p$ and $k$ are decay constants in the dielectric and metal, and $w$ is the insulator width.

In the proposed devices, we examine two hybrid resonator geometries: one with a circular silicon core and one with a square silicon core. Both operate in whispering-gallery–type regimes, where light circulates along the cavity perimeter with confinement assisted by total internal reflection \cite{chen2021recent}. The central silicon inclusion serves as a high-index perturbation that reshapes the modal field and boundary conditions, thereby modifying the coupling and loss pathways relative to a core-less cavity.

\begin{figure}[htbp]
\centering\includegraphics[width=13cm]{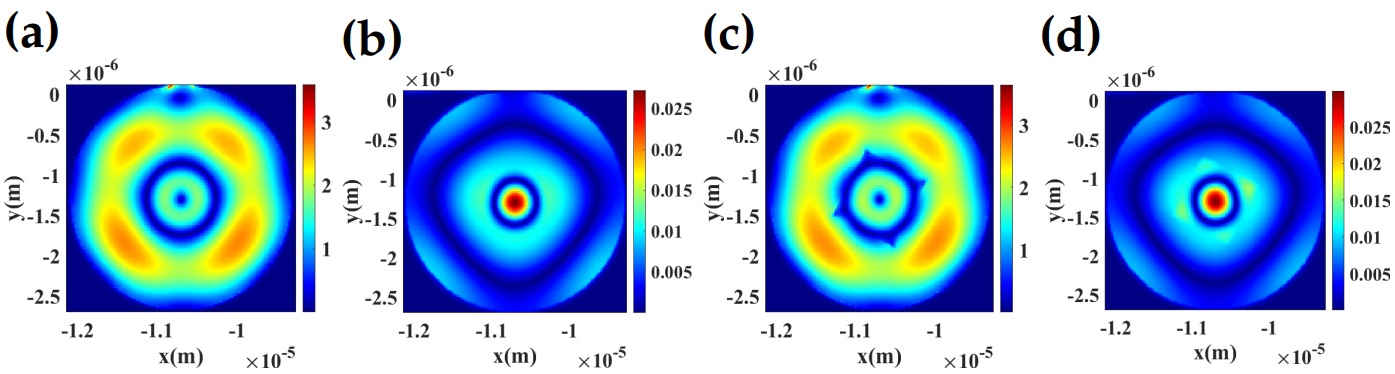}
\caption{Transmission spectra at $n_a=1.33$ for (a) cavity without core, (b) cavity with Ag core, and (c) cavity with Si core; (d) Si-core spectrum shown alone for clarity.}
\label{fig:2}
\end{figure}

Figure~\ref{fig:2}(a–c) compares simulated transmission spectra for three configurations at $n_a=1.33$: (a) no core, (b) silver (Ag) core, and (c) silicon (Si) core; panel (d) shows the Si-core spectrum alone for clarity. In our model, introducing a core produces a pronounced resonance that is absent in the core-less case. Between the two materials, the Si-core cavity exhibits a narrower linewidth than the Ag-core cavity, consistent with the lower absorptive loss of Si in the simulated wavelength range. The reduced linewidth implies a higher apparent quality factor $Q$ (extracted from Lorentzian fits), which in turn supports an improved figure of merit (FOM).

\begin{figure}[htbp]
\centering\includegraphics[width=13cm]{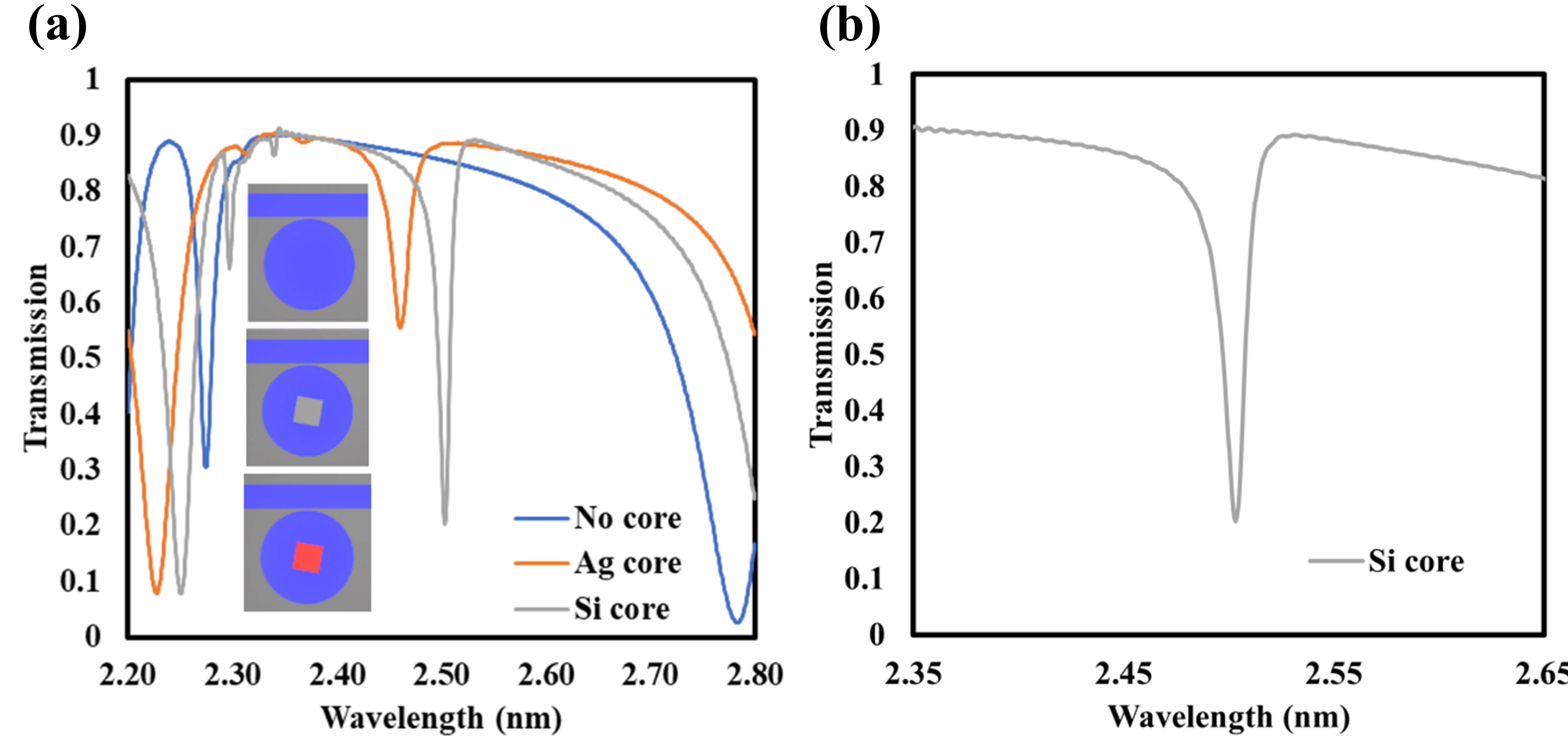}
\caption{Field maps for (a) electric and (b) magnetic fields in the circular-core resonator at $2.48\,\mu\mathrm{m}$, and (c) electric and (d) magnetic fields in the square-core resonator at $2.50\,\mu\mathrm{m}$.}
\label{fig:3}
\end{figure}

To further interpret the resonances, we visualize the electric- and magnetic-field distributions inside the cavity. Figure~\ref{fig:3}(a–d) shows the fields for a circular-core resonator at $2.48\,\mu\mathrm{m}$ and a square-core resonator at $2.50\,\mu\mathrm{m}$. As seen in Figure~\ref{fig:3}(b,d), elevated magnetic-field amplitudes occur near and within the silicon inclusion for both geometries, indicative of strong modal circulation and enhanced overlap with the sensing region. Compared with the Ag-core case, the Si core’s lower material loss yields tighter confinement and a smaller radiative/absorptive linewidth, consistent with the higher $Q$ and FOM reported in Section~\ref{sec:Evaluation}.

We assessed performance using the Lumerical FDTD solver for both core shapes under identical outer geometry and coupling conditions. Unless otherwise noted, the waveguide/cavity thickness $w$, the gap $g$ between the bus waveguide and ring, the ring outer radius $r$, and the silicon-core radius (or half-width) $r_c$ were held fixed across comparisons; only the core shape (circular vs.\ square) was varied. Two optimization strategies were considered: (i) manual tuning to establish a baseline and (ii) a GA search to navigate the multi-parameter space more systematically. Performance metrics follow standard definitions:
\begin{equation}
S=\frac{\Delta\lambda}{\Delta n}, \qquad
Q=\frac{\lambda_{\mathrm{res}}}{\mathrm{FWHM}}, \qquad
\mathrm{FOM}=\frac{S}{\mathrm{FWHM}},
\tag{5–7}
\end{equation}
where $S$ is the bulk RI sensitivity (nm/RIU), $\lambda_{\mathrm{res}}$ is the resonance wavelength, and FWHM is the full width at half maximum.

Under these assumptions, the simulations indicate that both the core material and its geometry influence the balance between analyte–field overlap and optical loss. The silicon core consistently yields narrower resonances (higher $Q$) than a metallic core at the same geometry, and the square versus circular cross-sections produce modest but reproducible differences in $Q$, FWHM, and FOM. These effects are examined in detail in the next section, with emphasis on hemoglobin-induced RI changes as a representative biochemical sensing scenario.

% -------------------- Rationale + Implementation (matches script) --------------------
\section{Genetic algorithm rationale and implementation}

\subsection{Rationale}
Optimizing the resonator requires searching a nonconvex space of geometric variables with many local optima. In preliminary trials, gradient-based solvers stalled at local minima. We therefore used a genetic algorithm (GA) to maintain a small population of candidate designs and apply selection, crossover, and mutation to escape local traps.

\subsection{Implementation}
We tuned three variables of the inner inclusion—the in-plane semi-axes $r_x,r_y$ (nm) and rotation angle $\theta$ (deg). Each candidate is $[r_x,r_y,\theta]$ with $r_x,r_y\in[400,700]$\,nm and $\theta\in[0^\circ,180^\circ]$.
Electromagnetic responses were computed in \emph{Lumerical FDTD Solutions} (v222) via the Python API (Lumapi) using a 2D domain, uniform mesh ($\Delta x=\Delta y=20$\,nm), simulation time 3000\,fs, and a mode source injected along the $x$-axis. Transmission was sampled by a linear-$y$ power monitor over 2.4–2.8\,$\mu$m with 200 points. The stack comprises a silver substrate (Palik, 1–10\,$\mu$m), a user-defined \texttt{Core} medium with real index $n$ and $k{=}0$ for the waveguide region, and a silicon (Palik) elliptical inclusion.

Fitness is the refractive-index sensitivity
\begin{equation}
S = \frac{\Delta\lambda_{\max}}{\Delta n}, \qquad \Delta n = 0.03~\text{RIU}.
\end{equation}
where $\Delta\lambda_{\max}$ is the largest separation between corresponding transmission dips for $n\in\{1.33,1.36\}$. Dips are identified where $T(\lambda)<0.2$, and adjacent samples are de-duplicated by enforcing a spacing of at least 15 spectral indices. To suppress spurious values, candidates with $S>2\times10^{-6}$\,m/RIU are assigned zero fitness.

The GA used a population of $2$ for $2$ generations. Parents were chosen by tournament selection (size $\min\{5,N\}$). Crossover operated at the \emph{parent level} (each child is copied from one parent with probability $0.5$; no per-gene mixing). Mutation was applied per individual with probability $0.1$, adding Gaussian jitter with standard deviation $0.1$ to each of $r_x$, $r_y$, and $\theta$. The script reports the best fitness from the last evaluated population and plots spectra for that candidate.

\begin{table}[ht]
\centering
\caption{Genetic algorithm settings used in the script.}
\label{tab:ga_parameters}
\renewcommand{\arraystretch}{1.12}
\small
\begin{tabular}{l l}
\hline
\textbf{Component} & \textbf{Setting} \\
\hline
Design variables & $r_x,r_y \in [400,700]$\,nm;\; $\theta \in [0^\circ,180^\circ]$ \\
Wavelength span & 2.4--2.8\,$\mu$m (200 samples) \\
Solver & 2D FDTD; uniform mesh $\Delta x=\Delta y=20$\,nm; Lumapi (v222) \\
Materials & Ag substrate (Palik);\; \texttt{Core} ($n$, $k{=}0$);\; Si (Palik) ellipse \\
Fitness & $\Delta\lambda_{\max}/0.03$ from dips with $T<0.2$; $\ge$15-sample spacing \\
Population size & 2 \\
Generations & 2 \\
Selection & Tournament (size $\min\{5,N\}$) \\
Crossover & Parent copy, $p{=}0.5$ (no per-gene mixing) \\
Mutation & Per-individual, $p{=}0.1$; Gaussian jitter $\sigma{=}0.1$ on each gene \\
Termination & Fixed at 2 generations; $S>2\times10^{-6}$\,m/RIU $\rightarrow$ fitness $0$ \\
\hline
\end{tabular}
\end{table}

\begin{algorithm}
\caption{Genetic algorithm loop (as implemented)}
\label{alg:ga}
\begin{algorithmic}[1]
\Procedure{Evaluate}{$c=[r_x,r_y,\theta]$}
  \State Run 2D FDTD via Lumapi at $n\in\{1.33,1.36\}$; sample 2.4--2.8\,$\mu$m (200 pts)
  \State From $T(\lambda)$, find dips with $T<0.2$; enforce $\ge$15-sample spacing
  \State $\Delta\lambda_{\max}\gets$ largest separation between matched dips
  \State $S\gets \Delta\lambda_{\max}/0.03$; \Return $0$ if $S>2\times10^{-6}$ m/RIU else $S$
\EndProcedure
\Procedure{Optimize}{}
  \State $N\gets2$;\; $G\gets2$;\; $P\gets$ \textsc{InitializePopulation}$(N)$ within bounds
  \For{$g=1$ \textbf{to} $G$}
    \State $f \gets [\textsc{Evaluate}(c): c\in P]$
    \State $s\gets \min(5,N)$;\; $P_s\gets$ \textsc{TournamentSelect}$(P,f,s)$
    \State $O\gets$ \textsc{ParentCopyCrossover}$(P_s,\,p=0.5)$ \Comment{no per-gene mixing}
    \State $O\gets$ \textsc{MutatePerIndividual}$(O,\,\text{rate}=0.1,\,\sigma=0.1)$
    \State $P\gets O$
  \EndFor
  \State \Return $\arg\max_c f$ from last evaluation;\; \textsc{PlotSpectra}$(c)$
\EndProcedure
\end{algorithmic}
\end{algorithm}

\noindent\textbf{Practical considerations.} With a 20\,nm mesh, the mutation jitter ($\sigma=0.1$) is effectively sub-grid; larger perturbations or per-gene mutation may improve exploration. Because the crossover operator copies whole parents, no per-gene mixing occurs; a uniform per-gene crossover would increase diversity. Finally, the script reports the best-scoring candidate from the last evaluated population rather than re-evaluating the final population.

% -------------------------------------------------------------------

\section{Evaluation of Biosensing Performance}

\begin{figure}[htbp]
\centering\includegraphics[width=13cm]{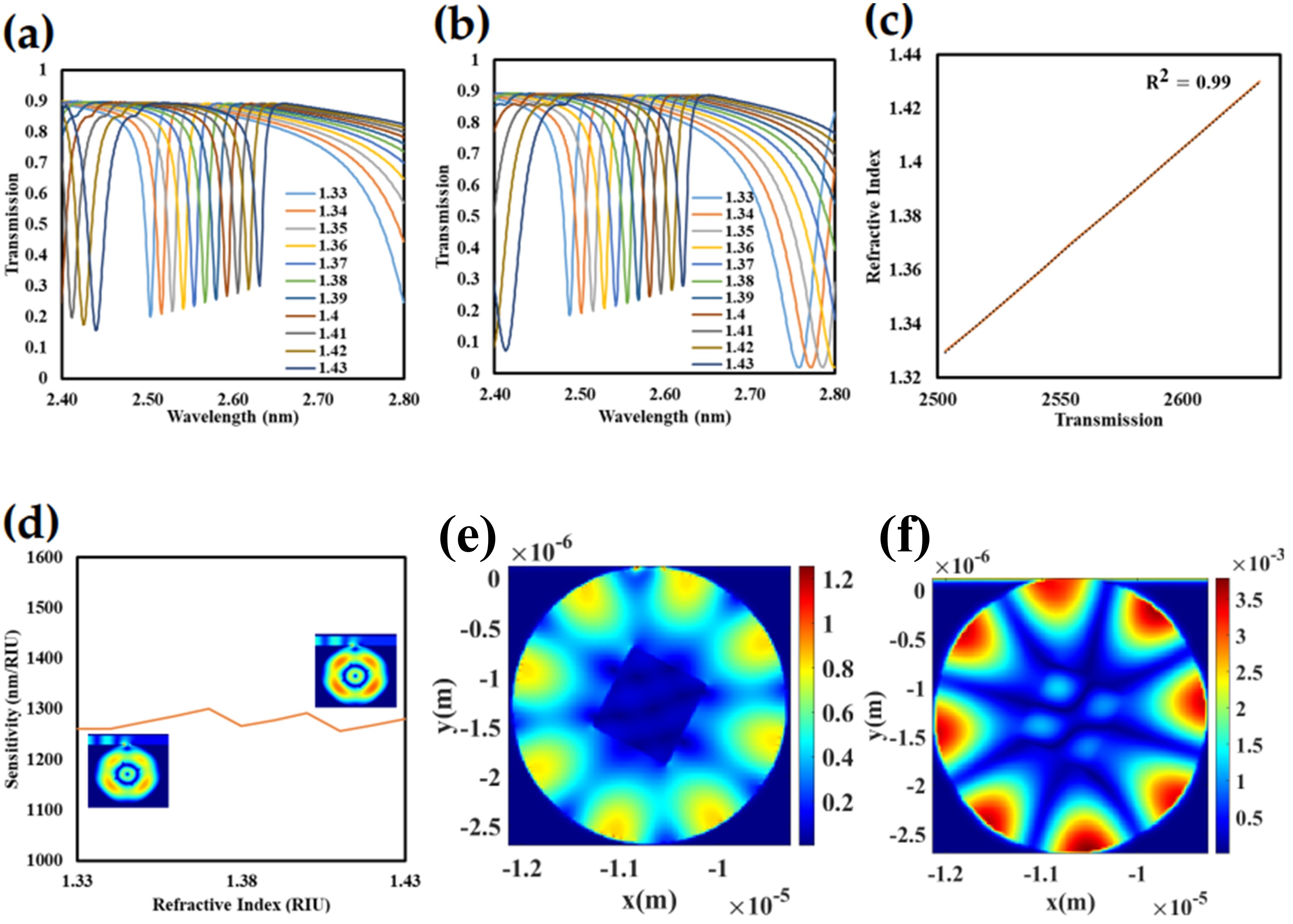}
\caption{(a) Transmission spectra for square and (b) circular cores at varying indices. (c) Linearity graph of the square core. (d) Sensitivity variations in the square core from \( n_a=1.33 \) to \( n_a=1.43 \). (e, f) Electric and magnetic field distributions in the hybrid resonator with a square core in Whispering Gallery Mode.}
\label{fig:4}
\end{figure}

Hemoglobin (Hgb) is a clinically relevant analyte for disorders such as polycythemia vera, hereditary anemia, and hematuria; its concentration modulates the refractive index (RI) of the host medium and can therefore be probed optically \cite{patel2020graphene}. In our simulations, transmission spectra were computed for Hgb concentrations from $10$ to $40\,\mathrm{g/L}$, corresponding to bulk RI values in the range $n\!\approx\!1.33$–$1.43$. Figure~\ref{fig:6} shows three-dimensional schematic of the proposed sensor chip. The right panel shows the full device; the dashed box indicates the region magnified in the left insets. The insets compare two core geometries—square and circular pillars (red)—embedded within the sensing cavity (blue). Ribbon models depict hemoglobin molecules rendered approximately to molecular scale ($\sim$5--6\,nm) for size reference. Figure~\ref{fig:4}(a) shows the spectra for the square core, and Figure~\ref{fig:4}(b) shows the corresponding response for the circular core. We first established a manually tuned baseline. The square core performed best at $1300\,\mathrm{nm}\times 1050\,\mathrm{nm}$, yielding a sensitivity $S=845\,\mathrm{nm/RIU}$, full width at half maximum $\mathrm{FWHM}=16.44\,\mathrm{nm}$, figure of merit $\mathrm{FOM}=51.39\,\mathrm{RIU^{-1}}$, and quality factor $Q=162$. The circular core optimized to a radius of $400\,\mathrm{nm}$ with $S=1218\,\mathrm{nm/RIU}$, $\mathrm{FWHM}=10.78\,\mathrm{nm}$, $\mathrm{FOM}=112\,\mathrm{RIU^{-1}}$, and $Q=244$. To explore the multi-parameter design space more systematically, we then coupled a GA to the FDTD solver. GA refinement for the square core converged to $906\,\mathrm{nm}\times 854.5\,\mathrm{nm}$ and improved the metrics to $S=1276\,\mathrm{nm/RIU}$, $\mathrm{FWHM}=11.89\,\mathrm{nm}$, $\mathrm{FOM}=107\,\mathrm{RIU^{-1}}$, and $Q=221$. For the circular core, the GA selected a radius of $468\,\mathrm{nm}$, giving $S=1386\,\mathrm{nm/RIU}$, $\mathrm{FWHM}=13.13\,\mathrm{nm}$, $\mathrm{FOM}=96.95\,\mathrm{RIU^{-1}}$, and $Q=199$. Across the Hgb sweep, the resonance shifted from $\lambda_{\mathrm{res}}=2503.89\,\mathrm{nm}$ at $10\,\mathrm{g/L}$ to $2631.50\,\mathrm{nm}$ at $40\,\mathrm{g/L}$, consistent with the observed sensitivities and narrow linewidths. A direct post-GA comparison highlights a modest trade-off: the circular core provides the larger $S$, whereas the square core maintains a slightly higher $Q$ and $\mathrm{FOM}$. Linear fits of $\lambda_{\mathrm{res}}$ versus $n$ for both geometries exhibit near-unity coefficients of determination ($R^2=0.99$), as summarized in Figure~\ref{fig:4}(c), indicating robust linear transduction over the tested RI range. The sensitivity trend in Figure~\ref{fig:4}(d) averages $S\approx 1276\,\mathrm{nm/RIU}$ with small fluctuations attributed to numerical tolerances; a mild increase from $1261\,\mathrm{nm/RIU}$ at $n=1.33$ to $1280\,\mathrm{nm/RIU}$ at $n=1.43$ is observed. Field maps in Figure~\ref{fig:4}(e,f) visualize whispering-gallery–type confinement for both cores, with stronger azimuthal circulation evident in the square core, consistent with its higher measured $Q$.

\begin{figure}[htbp]
\centering\includegraphics[width=13cm]{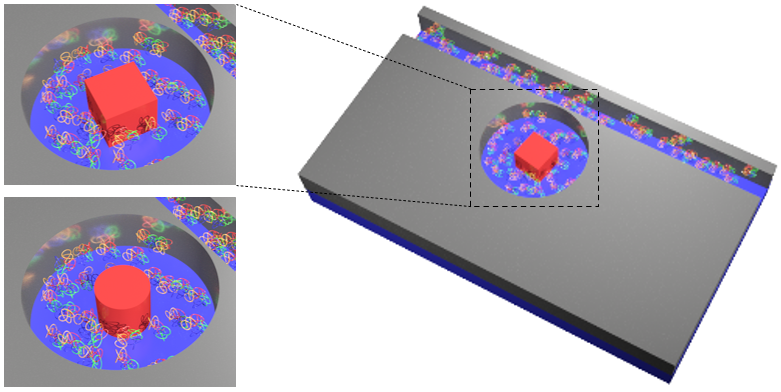}
\caption{3D schematic of the sensor with zoomed inset; insets compare square and circular cores with hemoglobin molecules ($\sim$5--6\,nm).}
\label{fig:7}
\end{figure}

Table~\ref{Tab. 1} compares representative refractive-index biosensors based on dielectric, plasmonic, and hybrid waveguide resonators, including the GA-optimized hybrid cavity reported here. Relative to prior designs, the present device achieves a favorable combination of sensitivity ($S$), FWHM, quality factor ($Q$), and $\mathrm{FOM}=S/\mathrm{FWHM}$. The reduced linewidth improves spectral discriminability, and together with the observed $Q$ indicates efficient energy storage. These characteristics support use in label-free biochemical assays, environmental monitoring, and food-quality analysis.

\begin{table}[htbp]
  \centering
  \caption{\bf Assessment of the Performance in Sensing}
  \small % Reducing the font size
  \begin{tabular}{p{2cm}p{2cm}p{2cm}p{2cm}p{2cm}p{2cm}} % Adjusted column widths
    \hline
    \textbf{Structure} & \textbf{Sensitivity (nm/RIU)} & \textbf{FWHM} & \textbf{FOM (RIU\(^{-1}\))} & \textbf{Q-factor} & \textbf{Ref.} \\
  \hline
    RC & 700 & - & 21.9 & - & \cite{kazanskiy2021numerical} \\
    HRR & 1609 & - & - & - & \cite{kumar2023nanophotonic} \\
    RR & 300 & 7.4 & 36.6 & 201.6 & \cite{butt2019sensitivity} \\
    RR & 1100 & 5 & 200 & - & \cite{khodadadi2020theoretical} \\
    HRR & 1000 & 3.48 & 287.35 & 441.05 & \cite{butt2020modal} \\
    HRR & 534 & - & 89 & - & \cite{kumari2022parametric} \\
    HRR & 555 & 25.8 & 154 & 434 & \cite{kumari2022hybrid} \\
    RC & 1250 & - & 88.68 & - & \cite{hu2021sensor} \\
    HRC & 1276 & 11.89 & 107 & 221 & This Work \\
    \bottomrule
      \hline
  \end{tabular}%
  \label{Tab. 1}%
\end{table}

\section{Conclusion}
The optimized hybrid plasmonic resonator provides a compact, CMOS-compatible platform for refractive-index biosensing. GA tuning of the core geometry increases analyte–field overlap while preserving a narrow resonance. The device exhibits an average sensitivity of $S=1276\,\mathrm{nm/RIU}$, a full width at half maximum of $\Gamma=11.89\,\mathrm{nm}$, a quality factor of $Q=221$, and a FOM $\mathrm{FOM}=S/\Gamma=107\,\mathrm{RIU}^{-1}$. The small $\Gamma$ improves spectral resolvability for precise analyte discrimination, while the observed $Q$ indicates efficient energy storage and a favorable signal-to-noise ratio under wavelength or intensity interrogation. Taken together, these metrics place the sensor among competitive hybrid plasmonic implementations and support applications in label-free biomedical diagnostics, environmental monitoring, and food-quality control.

% \disclosures 
\subsection*{Disclosures}
The authors declare that there are no financial interests, commercial affiliations, or other potential conflicts of interest that could have influenced the objectivity of this research or the writing of this paper.

\subsection*{Code, Data, and Materials Availability}
The analysis and optimization code used in this study is publicly available at GitHub: \url{https://github.com/Alireza36991/OE-Biosensor}.

%%%%% References %%%%%

\bibliography{report}   % bibliography data in report.bib
\bibliographystyle{spiejour}   % makes bibtex use spiejour.bst

%%%%% Biographies of authors %%%%%

\vspace{2ex}\noindent\textbf{Alireza Khalilian} is a Ph.D. candidate in the Department of Electrical and Computer Engineering at the University of Michigan–Dearborn. He received his M.S. in Electrical Engineering from Kyungpook National University in 2017. He has authored and coauthored several journal articles and conference papers. His current research focuses on metasurfaces, metalens design, hybrid plasmonic–dielectric resonators, integrated photonics, and optoelectronic systems for imaging and polarization detection. He is a member of SPIE and Optica.

\vspace{2ex}\noindent\textbf{Hima Nikafshan Rad} is a Ph.D. candidate in the School of Information and Communication Technology at Griffith University, Australia, and a member of the AI for Life Laboratory. Her research centers on machine learning and deep learning for biomedical and energy systems, with recent work on multi-omic integration for amyotrophic lateral sclerosis (ALS) diagnosis and on low-carbon energy system analysis. She has industry experience as a machine-learning researcher with GenieUs Genomics. She has authored peer-reviewed articles spanning biomedical AI and energy/exergy optimization.

\listoffigures
\listoftables

\end{spacing}
\end{document}